\RequirePackage{fix-cm}

\documentclass[smallextended]{svjour3}

\smartqed
\usepackage{moreverb}
\usepackage[colorlinks,bookmarksopen,bookmarksnumbered,citecolor=red,urlcolor=red]{hyperref}
\usepackage{amssymb,amsmath}
\usepackage{graphicx}
\usepackage{subfigure}
\usepackage{bm,bbm,booktabs,units,multirow}

\usepackage[ruled]{algorithm}
\usepackage{algpseudocode}

\graphicspath{{figures/}{./CelsoFaccio/}}

\begin{document}

\title{Stochastic particle packing with specified granulometry and porosity}
\titlerunning{Stochastic particle packing with specified granulometry and porosity}        

\author{Alejandro C. Frery\and \mbox{Lorena Rivarola-Duarte}\and \mbox{Viviane Carrilho Le\~ao Ramos}\and \mbox{Adeildo Soares Ramos Junior}\and \mbox{William Wagner Matos Lira}}
\authorrunning{A.C.\ Frery et al.} 

\institute{A.C.\ Frery, V.C.L.\ Ramos, A.S.\ Ramos Junior, W.W.M.\ Lira\at
Laborat\'orio de Computa\c c\~ao Cient\'ifica e Visualiza\c c\~ao\\
Universidade Federal de Alagoas\\
Campus A.\ C.\ Sim\~oes\\
Av. Lourival Melo Mota, s/n,
Tabuleiro do Martins\\
57072-970 Macei\'o, AL -- Brazil\and
L.\ Rivarola-Duarte\at
Universidade Federal de Minas Gerais}

\date{Received: date / Accepted: date}

\maketitle

\begin{abstract}
This work presents a technique for particle size generation and placement in arbitrary closed domains.
Its main application is the simulation of granular media described by disks. 
Particle size generation is based on the statistical analysis of granulometric curves which are used as empirical cumulative distribution functions to sample from mixtures of uniform distributions.
The desired porosity is attained by selecting a certain number of particles, and their placement is performed by a stochastic point process.
We present an application analyzing different types of sand and clay, where we model the grain size with the gamma, lognormal, Weibull and hyperbolic distributions.
The parameters from the resulting best fit are used to generate samples from the theoretical distribution, which are used for filling a finite-size area with non-overlapping disks deployed by a Simple Sequential Inhibition stochastic point process.
Such filled areas are relevant as plausible inputs for assessing Discrete Element Method and similar techniques.
\keywords{granular media \and simulation\and particulate systems\and particle generation\and point processess}
\end{abstract}

\maketitle

\section{Introduction}

The Discrete Element Method (DEM) has become a powerful tool in the numerical simulation of engineering problems involving discontinuous media.
Among the problems where the method has been applied, we may cite fragmentation, fracture, impact, and collision phenomena, in addition to those directly related to soil modeling as part of studies of geomechanical problems~\cite{cundall88,cundall79,zarate01,zarate04}.

%

DEMs are based on media discretization into finite sets of particles.
More often than not, this discretization must satisfy some requirements.
In order to achieve significant results, a good representation of the studied media is needed and, therefore, the particle generation must meet the granulometric curves of soil and should be as close as possible to the specified porosity.
A general procedure would be then comprised of two stages, namely, (i)~particle size generation, and (ii)~particle placement.

The measurement of particle size is an activity that is common to diverse disciplines: archaeology, fuel technology, medicine, geology, sedimentology, civil engineering, pharmacology, etc.
The common objective is to determine the overall size distribution of a collection of particles~\cite{Fieller1992}.
Particle-size distribution (PSD) is fundamental for characterizing construction materials, soil mechanics, soil physics, sediment-flux in rivers, and others.
In soil science, it is typically presented as the percentage of the total mass of soil occupied by a given size fraction.
Determination of the soil PSD is not a trivial task because of the heterogeneity of the shape and density of particles~\cite{Eshel2004}.
Classical techniques used to determine the PSD, like sieving and sedimentation, are point-wise and, thus, require an interpolation to obtain the complete PSD curve. 
The transformation of discrete points into continuous functions can be made by mathematical models~\cite{daSilva2004}, where the normal, lognormal and Weibull distributions are especially prevalent~\cite{Fieller1992}.

The PSD in soils is frequently assumed to be approximately lognormal~\cite{Buchan1989,Shirazi1984}, but there are soils which present bimodal PSDs~\cite{Walker1986}.
A bimodal lognormal Gaussian distribution is presented in~\cite{Shiozawa1991} for the characterization of various soil samples.
It consists of a weighted sum of two distributions: primary minerals (sand and silt), and secondary minerals (clay), each described by a Gaussian law.

Besides the lognormal distributions, other models have been proposed, including one based on the water retention curve~\cite{Fredlund2000}, the Gompertz model~\cite{Nemes1999}, the fragmentation model~\cite{Bittelli1999} and estimating the PSD from limited soil texture data~\cite{Skaggs2001}.
Furthermore, the modelling of PSD by means of the fractal mass distribution is presented in~\cite{Martin1998}. 
Combining some well-founded theoretical results from fractal geometry, the model allows to simulate the PSD of a given soil and its characterization by means of the entropy dimension.

Although statisticians have ocasionally examined particular problems with the estimation of PSD~\cite{Fisher1923}, it was not until the work of~\cite{Barndorff-Nielsen1977} that a coherent statistical approach was formulated.
That paper introduced the log-hyperbolic distribution as a suitable model for particle sizes.
However, the considerable computational difficult involved in fitting this model has prevented it from achieving widespread use~\cite{Fieller1992}.

The comparison of mathematical models for fitting PSD curves in soil science has been performed in a few works~\cite{Buchan1993,daSilva2004,Hwang2002,Rousseva1997}.
On the studies of~\cite{Hwang2002}, seven parametric models were tested -- the majority lognormal models and growing curves type -- including a unimodal and six bimodal models.
They compare five lognormal models with one, two and three parameters~\cite{Buchan1993}, the Gompertz model with four parameters~\cite{Nemes1999} and the Fredlund model with four parameters~\cite{Fredlund2000} using 1387 Corean soil samples.
Four comparison techniques were considered to define the best model: the coefficient of determination ($R^2$), the $F$ statistic, the $C_p$ statistic of Mallows~\cite{Mallows1973} and the Akaike's information criterion (AIC).
They conclude that the Fredlund model presented the best fit for the majority of the soils studied, with increasing performance with the increase of clay content. 
Furthermore, they showed that texture could affect the performance of the PSD models.
This work represents an important contribution to the model comparison for fitting granulometric data, but another models, potencially adaptable to this finality were not studied.
Moreover, the authors compared models with different number of parameters; this could be a problem because it can benefit those more degrees of freedom~\cite{daSilva2004}.

The work done by~\cite{daSilva2004} tested and compared fourteen different models with feasibility to fit the cumulative PSD curve based on four measured points.
The parameter used to compare the models was the sum of the square errors between the measured and calculated values.
They concluded that the most recommendable models to fit PSD curves were: Skaggs model~\cite{Skaggs2001}, Weibull model \cite{Weibull1951} and Morgan model \cite{Morgan1975}, all of them with three parameters.

In this work we transform data from granulometric curves into unevenly spaced histograms, and then sample from this empirical distribution.
The sampled data is explained by means of four distributions, namely the hyperbolic, gamma, log-normal and Weibull laws, and the best fit is associated to the original granulometric curve.
We then obtain an arbitrary number of radii sampling from the chosen distribution, and we place these particles in an arbitrary closed region of the plane or the space, using a stochastic point process.
The number of particles to be used is iteratively determined in order to satisfy a predefined porosity.

Many procedures have been presented for particle placement with well-known tractable shapes, especially disks and balls (spheres) for the two or three- dimensional cases, respectively.
These procedures are generally divided into two distinct groups: geometric and dynamic algorithms.

Dynamic algorithms use mechanical procedures that reproduce external actions to achieve an initial set of particles.
A typical approach of these algorithms consists of creating a regular array of particles and them randomly disorder it through the action of gravitational forces~\cite{feng03}.
Other approaches require the use of DEM simulation procedures in order to achieve the initial configuration.

These dynamic approaches have many advantages, including the possibility of filling any arbitrary-geometrical domain with elements of predefined, irregular sizes and, still, controlling the desired porosity.
An overview of typical dynamic techniques can be found in~\cite{Cui03,feng03}.
However, in most cases, these dynamic algorithms require checking of contacts between particles to achieve the desired result, making them significantly slower in terms of computational cost.

Another category of techniques is based on geometric algorithms.
These algorithms are characterized by the use of purely geometric concepts for generating particles in a given domain, without the need of dynamic simulations to represent their movement and interactions.
This often reduces significantly the processing time.
The literature of granular media offers several techniques for random generation of disks or spheres with different diameters.
One of the main strategies consists on establishing the position and the size of particles from random numbers.
If there is overlap, a new random location is performed with the same fixed size~\cite{PFC2D,Lin1997Geotechnique}.

Other geometric techniques are based on particle generation from an initial triangular or tetrahedrons elements mesh, and placing disks or spheres in the interior of these elements~\cite{Cui03}.
Several other geometric strategies can be found in the literature, including those that use the concept of boundary contraction, where the particle locations are calculated from the previous inclusion of other particles with a pre-established diameter~\cite{Bagi2005GranularMatter,Labra2008CommunicationsinNumericalMethodsinEngineering,Lohner2004InternationalJournalforNumericalMethodsinEngineering}.

The class of Random Sequential Adsorption -- RSA processes describes the deposition of particles. Variations include the dimension and other properties of the substrate, the shape of the particles and how they interact among them and with the media~\cite{DynamicsNonequilibriumDepositionPrivman}.
An important characteristic of our work is that we use random particle sizes. 
Quoting Cadilhe et al~\cite{RandomSequentialAdsorption}:
\begin{quote}
Versions of the continuum RSA model with deposition of mixtures of particles of different sizes and shapes have only been studied for the simplest cases of two particle sizes and identical shapes. [\dots] In spite of all the research work thus far, the field remains widely open to new research efforts.
\end{quote}

Dynamic algorithms grant the simulation of samples which are in geostatic balance with spatial homogeneity, and either specified granulometry or prosity, at the expense of high computational cost.
Geometric algorithms are much faster than the former, but they do not grant geostatic balance.
The latter are, thus, more adequate in, for instance, Monte Carlo techniques.
Moreover, traditional geometric algorithms have difficulties in reaching answers that meet the specified granulometric curves.

\begin{figure}[hbt]
\centering
\includegraphics[bb=315 132 1040 206,clip,width=3.4\linewidth]{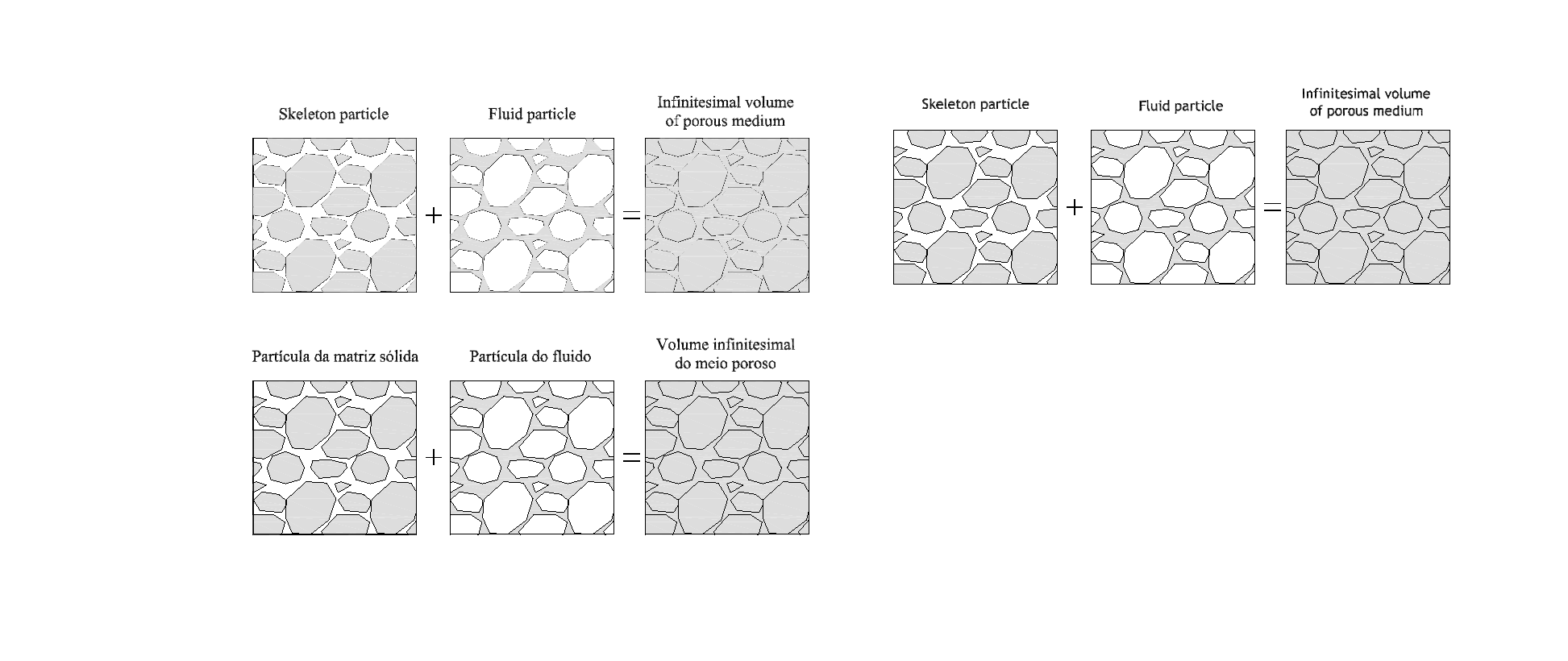}
\caption{Fully saturated porous media (From~\cite{PoromechanicsCoussy})}
\label{fig:FullySaturatedPorous}
\end{figure}

This work presents a geometric technique for particle size generation and placement in arbitrary closed domains.
Its main application is the simulation of granular media described by disks which can be extended to spheres. 
Particle size generation is based on the statistical analysis of granulometric curves.
These curves are treated as empirical cumulative distribution functions, histograms are derived from them, and each bin is used as the density of a uniform random variable.
Samples from these uniform laws are then obtained.
The desired porosity is attained by selecting a certain number of particles, and their placement is performed by a stochastic point process.
This proposal grants both porosity and granulometric properties with spatial homogeneity, i.e, the system is in global equilibrium as in fully saturated porous media (see Figure~\ref{fig:FullySaturatedPorous}).
On the one hand, by construction, the results adhere to the statistical properties of the observed data. 
On the other hand, the porosity we measured is an underestimate of the true property; nevertheless, the method accepts and produces any value of porosity, so it is up to th

The proposed strategy is comprised of the following steps:
\begin{enumerate} 
 \item Particle size generation:
\begin{enumerate}
  \item A granulometric curve and the porosity are used as input (section~\ref{sec:granulometriccurve}).
  \item A histogram is derived from this curve, and samples are drawn from the empirical distribution provided by the histogram (section~\ref{sec:histogram}).
  \item The samples are used to estimate the parameters of a set of distributions (section~\ref{sec:models}).
  \item The estimated distribution that best fits the original data is adopted (section~\ref{sec:bestfit}).
  \item An approximate number of particles that provide the desired porosity is calculated (section~\ref{sec:number}).
  \item The radii of the particles are samples of independent identically distributed random variables obeying the adopted distribution (section~\ref{sec:radii}).
  \end{enumerate}
\item The particles are placed by a Simple Sequential Inhibition -- SSI process on the arbitrary closed region  (section~\ref{sec:placement}). The number of particles is determined iteratively as to produce an approximation of the desired porosity.
\end{enumerate}

The result of these steps is a plausible model of particle distribution in granular media, obtained with low computational cost due to its geometric nature.
Algorithm~\ref{algo:sequential} presents, in pseudocode, an efficient implementation of this procedure.
We show that this procedure is able to simulate soils of low porosity, which are usually obtained by dynamical techniques that are much more time consuming.
Such models can be used as input data for numerical simulations of problems with discontinuous media using the discrete element method.

The SSI process we employ can be seen as a RSA process with the following characteristics: the particles only interact by exclusion in a compact planar domain, i.e., there is neither attraction nor repulsion, no order is introduced in the process, i.e., the final state is jammed, particles arrive sequentially and there is no relaxation, i.e., the $(k+1)$th particle enters the sample only if the $k$th particle does.

This paper unfolds as follows.
Section~\ref{sec:methodology} presents the methodology of each stage of our proposal, including pseudocode algorithms.
Section~\ref{sec:results} discusses the main results, while section~\ref{sec:conclusion} presents the conclusions and future venues of research.

\section{Particle size analysis and generation}\label{sec:methodology}

\subsection{Granulometric curve and porosity}\label{sec:granulometriccurve}

A granulometric curve measures the percentage of the sample that falls into pre-established ranges of grain sizes.
This information is obtained by seiving and hydrometer analyses, for instance.
The results of such measures is provided in tabular form: $(d_i,c_i)_{1\leq i\leq D}$, being $c_i$ the (cumulative) proportion of particles whose diameter is less than $d_i$, and $D$ the number of diameters considered.

The porosity $\eta\in[0,1]$ is the ratio between the empty space in the sample to the total sample volume.
It is estimated through the determination of the bulk density of the porous sample, then determining the density of the skeletal material, and finally correlating density and volume.

The granulometric curve and the porosity are the only required input for the rest of the methodology.

\subsection{Histograms}\label{sec:histogram}

The granulometric curve is used to form a tractable and expressive histogram of particle sizes.
As presented in Figure~\ref{c1}, the diameters are not evenly spaced and, thus, the intervals are not equal.
Such unevenness is alleviated in the semilogarithmic scale, so our proposal uses such transformation.

Consider the original data $(d_i,c_i)_{1\leq i\leq D}$ and its representation in semilogarithmic scale $(\ell_i,c_i)_{1\leq i\leq D}$, where $\ell_i=\log d_i$ for every $1\leq i\leq D$.
The log-histogram of the data can be formed with the midpoints of pairs of contiguous log-diameters and the corresponding proportion of particles, i.e., $(m_i,p_i)_{1\leq i\leq D-1}$, where $m_i=(\ell_{i+1}+\ell_i)/2$ and $p_i=c_{i+1}-c_i$.
These log-histograms will be used to obtain more data from pseudorandom sampling, in order to obtain parametric models for the data.

Each dataset is, therefore, transformed into a log-histogram.
Instead of using the pairs $(m_i,p_i)_{1\leq i\leq D-1}$, we draw $N_i$ independent identically distributed samples from the Uniform distribution on $(\ell_i,\ell_{i+1})$, with $N_i=[kp_i]$ and $k$ a convenient number, for instance $k=1000$.
With this, we end with $\sum_{i=1}^{D-1} N_i = M$ pseudorandom diameters $\bm d=(d_1,\dots,d_M)$ which, if given as input, would produce the original granulometric curve $(d_i,c_i)_{1\leq i\leq D}$.
In this way, an arbitrary number of samples can be obtained from a single granulometric curve, leading to very precise parameter estimation.

These data $\bm d$ will be then used for computing estimates of parametric models able to describe the original data.

\subsection{Models}\label{sec:models}

\subsubsection{Hyperbolic distribution}

The unidimensional hyperbolic distribution is a continuous probability law characterized by the fact that the logarithm of the probability density function is a hyperbola.
It was introduced by~\cite{Barndorff-Nielsen1977} as a model for the log-size distribution of sand, based on the studies of~\cite{Bagnold1941} about aeolian sand deposits.

This distribution and its multidimensional extensions have many applications in a variety of fields: geology, astronomy, fluid mechanics and economics~\cite{Bagnold1980,Barndorff-Nielsen1978,Barndorff-Nielsen1981a,Barndorff-Nielsen1985}, to name a few.

One of the representations of the probability density function, presented in~\cite{Barndorff-Nielsen1983}, is:
$$ f(x;\alpha,\beta,\mu,\delta) = \frac{\sqrt{\alpha^2-\beta^2} \exp\{-\alpha\sqrt{\delta^2+(x-\mu)^2}+\beta(x-\mu)\}} {2\alpha\delta K_1(\delta\sqrt{\alpha^2-\beta^2}) },$$
where $x,\mu,\beta\in\mathbbm{R}$, $\delta,\alpha\in{\mathbbm{R}_+}$ such that $0\leq|\beta|<\alpha$, and $K_i$, $i=1,2,3$, denotes modified Bessel functions of the third kind.
The location parameter $\mu$ is the abscissa of the point of intersection between the linear asymptotes of the hyperbola, while $\delta$ is the scale parameter. 
The two parameters $\alpha$ and $\beta$ determine the shape, being $\alpha$  resposible for the steepness and $\beta$ for the skewness.
The distribution is symmetric for $\beta=0$.

If $X$ is a hyperbolic distributed variable, its expected value and mode are
$$E(X) = \mu + \frac{\delta\beta K_2(\delta\sqrt{\alpha^2-\beta^2})}{\sqrt{\alpha^2-\beta^2}K_1(\delta\sqrt{\alpha^2-\beta^2})}
\text{ and }\operatorname{Mode}(X) = \mu + \frac{\delta\beta}{\sqrt{\alpha^2-\beta^2}},$$
and its variance is
\begin{align*}
\operatorname{Var}(X) = &\frac{\delta K_2(\delta\sqrt{\alpha^2-\beta^2})}{\sqrt{\alpha^2-\beta^2}K_1(\delta\sqrt{\alpha^2-\beta^2})} + \\
&\frac{\beta^2\delta^2}{\sqrt{\alpha^2-\beta^2}} \left(\frac{K_3(\delta\sqrt{\alpha^2-\beta^2})}{K_1(\delta\sqrt{\alpha^2-\beta^2})} - \frac{K_2^2(\delta\sqrt{\alpha^2-\beta^2})}{K_1^2(\delta\sqrt{\alpha^2-\beta^2})}\right).
\end{align*}
Another representation of the density function of the hyperbolic distribution is:
$$ f(x;\pi,\zeta,\mu,\delta) = \frac{\exp\Bigl\{-\zeta\Bigl(\sqrt{1+\pi^2}\sqrt{1+(\frac{x-\mu}{\delta})^2} -\pi\frac{x-\mu}{\delta}\Bigr)\Bigr\} } {2\delta\sqrt{1+\pi^2} K_ 1(\zeta)},$$
where $x, \mu, \pi\in\mathbbm{R}$, and $\delta,\zeta\in{\mathbbm{R}_+}$. 
The parameter $\zeta$ is a measure of the degree of peakedness, and $\pi$ is a measure of the asymmetry of the distribution.
When $\pi=0$, the distribution is symmetric.
Together, they determine the shape of the distribution. $\mu$ and $\delta$ are parameters of location and scale.
This last parametrization is employed by the \verb+HyperbolicDist+ package of \texttt R~\cite{ManualR}, which also provides techniques for computing maximum likelihood estimates.

\subsubsection{Gamma distribution}

The gamma distribution is a continuous probability distribution of two parameters. 
It has been used in a wide range of disciplines: from climatology (in particular, to study rainfall; see \cite{Aksoy2000}), to material analysis \cite{Basu2009}, to name a few.

It has a scale parameter $\lambda$ and a shape parameter $\alpha$, both positive, in its density function:
$$
f(x;\alpha,\lambda) = \frac{x^{\alpha-1}e^{-{x}/{\lambda}}}{\lambda^\alpha\Gamma(\alpha)} {\mathbbm{1}}_{\mathbbm{R}_+}(x),
$$
where $\Gamma$ is the incomplete gamma function given by $ \Gamma(\alpha) = \int_{0}^{\infty} x^{\alpha-1}e^{-x}dx$.
Alternatively, it can be parameterized in terms of the rate $\beta = {1}/{\lambda}$.

If $X$ is a gamma distributed variable, its expected value, mode and variance are:
$$E(X) = \alpha\lambda, \quad \text{Mode}(X) = (\alpha-1)\lambda, \text{ and}\quad \text{Var}(X) = \alpha\lambda^2,
$$
respectively.

\subsubsection{Lognormal distribution}

The lognormal distribution is the distribution of any random variable whose logarithm is normally distributed.
It is widely used in physics, statistics, geology, economics, biology etc.
Its probability density function is
$$ f(x;\mu,\sigma) = \frac{1}{x\sigma\sqrt{2\pi}}\exp\Bigl\{-\frac{(\ln{x} - \mu)^2}{2\sigma^2}\Bigr\}{\mathbbm{1}}_{\mathbbm{R}_+} (x),$$
where $\mu\in\mathbbm R$ and $\sigma>0$.

If $X$ is a lognormally distributed variable, its expected value, mode and variance are
$$
E(X) = e^{\mu+{\sigma^2}/{2}},\quad
\text{Mode}(X) = e^{\mu-\sigma^2}, \text{ and}\quad
\text{Var}(X) = (e^{\sigma^2}-1)e^{2\mu+\sigma^2},
$$
respectively.

\subsubsection{Weibull distribution}

This law is often called the Rosin-Rammler distribution when used to describe the size distribution of particles \cite{Rosin1933}.
The probability density function of a Weibull random variable $X$ is
$$
f(x;\alpha,\lambda) = \alpha{\lambda^\alpha} {x^{\alpha -1}} e^{-(\lambda x)^\alpha} {\mathbbm{1}}_{\mathbbm{R}_+} (x),
$$
where $\alpha,\lambda > 0$ are the shape and the scale parameters, respectively.

If $X$ is a Weibull distributed variable, its expected value, mode and variance are:
$$
E(X) = \lambda\Gamma\Bigl(1+\frac{1}{\alpha}\Bigr),\quad
\text{Mode}(X) = \lambda\Bigl(\frac{\alpha-1}{\alpha}\Bigr)^\frac{1}{\alpha},$$
and
$$
\text{Var}(X) = \frac1{\lambda^2} \Bigl(
\Gamma\bigl(\frac{\alpha+2}{\alpha}\bigr)-
\Gamma^2\bigl(\frac{\alpha+1}{\alpha}\bigr)
\Bigr),
$$
respectively.

\subsection{Choice of the best model}\label{sec:bestfit}

Each sample is used to estimate the maximum likelihood parameters $\widehat\theta$ that index the four distributions $\mathcal D$ considered as candidate models, and then the $\chi^2$ goodness-of-fit test is applied.
In principle, the model with highest $p$-value should be chosen as the best statistical description and, in case of tiers, the one with least computational requirements should be the final choice.

Other goodness-of-fit test as, for instance, the Kolmogorov-Smirnov test, can be used, as well as other estimation procedures, e.g., moment or analogy estimators.

\subsection{Number of particles}\label{sec:number}

We will derive an initial guess for the number of particles to be generated in order to have a good approximation of the desired prosity.

Assume the closed area of interest, $W$, has size $\mu(W)$, which we want to fill with particles whose radii follow the parametric model $\mathcal D(\widehat \theta)$ chosen as previously discussed.
We also have the desired porosity $\eta$, estimated from the soil sample:
$$
\eta = \frac{V_e}{\mu(W)} = \frac{\mu(W)-V_p}{\mu(W)},
$$
where $V_e$ and $V_p$ are the empty and particles areas, respectively.
The area of a disk with radius $R$, distributed according to the law $\mathcal D(\widehat \theta)$ is a random variable given by $V = \pi R^2$.
Then, assuming that $N$ particles are placed within $W$,
$$
V_p = \pi \sum_{i=1}^N R_i^2.
$$
Using expected values, 
$$
\operatorname{E}(V_p) = \pi \sum_{i=1}^N \operatorname{E}(R_i^2) = \pi \sum_{i=1}^N [\operatorname{Var}(R_i) + \operatorname{E}^2(R_i)]= N\pi[\operatorname{Var}(R_i) + \operatorname{E}^2(R_i)],
$$
and both variance and expected value are given explicitely as functions of the estimated parameter $\widehat\theta$, so
$$
N = \frac{(1-\eta)\mu(W)}{\pi[\widehat{\operatorname{Var}}(R_i) + \widehat{\operatorname{E}}^2(R_i)]}.
$$
This number of particles $N$ is intended as a rough estimate of the needed number of radii required to fill $W$ with the desired porosity and granulometry.

\subsection{Generation of particles}\label{sec:radii}

Four distributions are considered in this work, namely the Hyperbolic, Gamma, Log-normal and Weibull laws.
Sampling from the two first requires specialized algorithms, samples from the third are the exponential transformation of normal deviates, and inversion is required in order to sample from the last one.
All these samples are produced calling functions available in the R platform~\cite{Crawley2007}.

Algorithm~\ref{algo:sequential} presents, in pseudocode, a sequential approach for obtaining a good approximation to the desired porosity $\eta$, while keeping the statistical properties of the particles.
In this algorithm, $K>1$ is a factor which controls the number of extra particles to be generated; we set $K=10$, a good compromise between the relatively low cost of generating outcomes versus the high cost of initializing the generator.

\begin{algorithm}[hbt]
\caption{Sequential sampling of radii for a desired porosity}\label{algo:sequential}
\begin{algorithmic}[1]
\Procedure{SequentialRadii}{$W$, $N$, $\eta$, $\mathcal D(\widehat\theta)$, $K$} \Comment{Region of interest, number of particles, porosity, model and factor}
  \State Obtain $\bm r = (r_1,\dots,r_{KN})$ samples from independent identically distributed random variables following the $\mathcal D(\widehat\theta)$ distribution
  \State Allocate the vector of coordinates $\bm c= ((x_1,y_1),\dots, (x_{KN},y_{KN})) $
  \State $i\gets 0$, $\eta_i \gets 1$ \Comment{Initialize counter and porosity}
  \While{$\eta_i>\eta$}
    \State condition $\gets \textrm{SSI}(W, i, \bm r, \bm c)$\label{line:call:SSI} \Comment{Defined on Algorithm~\ref{algo:SSI}}
    \If {condition = TRUE} 
	\State $i\gets i+1$ \Comment{Update counter}
	\State $\eta_i \gets \eta_{i-1}-\pi r_i^2 / \mu(W)$ \Comment{Update porosity}
      \Else
	\State break \Comment{No more particles will be placed}
    \EndIf
  \EndWhile
\If{$i=0$} 
  \State \Return FAIL \Comment{The first particle could not be placed in $W$}
\Else
  \State  \Return $((x_1,y_1,r_1),\dots,(x_i,y_i,r_i),\eta_i)$ \Comment{Returns coordinates, radii and current porosity}
\EndIf
\EndProcedure
\end{algorithmic}
\end{algorithm}

This algorithm places a number of particles obeying the $\mathcal D(\widehat\theta)$ distribution which approximates the desired porosity $\eta$ on the region of interest $W$.

It is noteworthy that the presented methodology can be immediately extended to compact volumes.

\subsection{Spatial distribution by the SSI process}\label{sec:placement}

Line~\ref{line:call:SSI} of Algorithm~\ref{algo:sequential} makes a call to the procedure presented in Algorithm~\ref{algo:SSI}, which implements the sampling from the simple sequential inhibition point processes.
Such processes are defined in terms of a window $W$, the number of points and the exclusion radii among them~\cite{BerthelsenMoller:PrimerSPP:2002}; in our case, the number of points is, instead of specified \textit{a priori}, controlled by the desired porosity $\eta$ on $W$.

The Simple Sequential Inhibition (SSI) point process for $n$ points on $W$ with exclusion radii $\bm r = (r_1,\dots,r_n)$ and maximum number of iterations $j_{\max}$ tries to place $n$ non-overlapping disks of radii $\bm r$ by testing sequentially each disk against the previous ones.
The first disk is placed uniformly on $W$, provided it does not surpass the boudaries.
Subsequent steps sample a point in $W$ uniformly and independently of previous disks, and verifies if there is no overlapping; if there is not, the disk is placed, otherwise up to $j_{\max}$ independent trials are made.
The algorithm stops when all the disks have been placed, or when the maximum number of iterations has been reached, whichever takes place first.
Our implementation of the SSI point process is presented in Algorithm~\ref{algo:SSI}.

\begin{algorithm}[hbt]
\caption{Simple sequential inhibition point process}\label{algo:SSI}
\begin{algorithmic}[1]
\Procedure{SSI}{$W, i, \bm r, \bm c$} \Comment{Region of interest, iteration, radii and coordinates}
  \State Define $j_{\max}$ \Comment{The maximum number of trials per particle}
  \State $j\gets 0$
  \While{$j<j_{\max}$}
    \State Sample $(x,y)$ uniformly on $W$
   \If {$\min_{1\leq k< i}\{d((x,y), (x_k,y_k))\geq r_i\}$} 
      \State $c(i) \gets (x,y)$ \Comment{Place particle $i$ at $(x,y)$}
      \State \Return {TRUE} \Comment{Success}
   \Else
      \State $j\gets j+1$
   \EndIf
  \EndWhile
  \State \Return {FALSE} \Comment{Unable to place the $i$-th particle}
\EndProcedure
\end{algorithmic}
\end{algorithm}

Notice that the order in which the particles are places is not altered.
In particular, if particle $i\leq 2$ does not fit after $j_{\max}$ trials, the algorithm returns the current state which consists of the previous $i-1$ particles.
Even if particle $i+1$ fits trivially the current state, it is not included.
Selecting particles by their size would imply independent identically distributed random variables model, and switching to a collection of correlated deviates: the order statistics.

Dense samples are, typically, harder to obtain than situations of high porosity.
In those cases it is likely that the denser the desired outcome the more results will have to be discarded until a sample with the desired property is obtained.

\section{Results}\label{sec:results}

This section presents results related to the two main steps already discussed, namely, particle size generation and placement.
The example used to illustrate this work is based on collected disturbed samples from the Barreiras formation and other fluvial deposits in Alagoas state, Brazil, in order to study their granulometric properties.

The granulometric information is obtained in tabular form: $(d_i,c_i)_{1\leq i\leq D}$, being $c_i$ the (cumulative) proportion of particles whose diameter is less than $d_i$, and $D$ the number of diameters considered.

\begin{figure}[hbt]
\centering
\begin{minipage}[c]{.3\linewidth}
\begin{tabular}{rr}
\toprule
Diameter [mm] & Cumulative \% \\ \midrule
4.800	& 100.00 \\
2.000	& 98.65 \\
1.200	& 83.81 \\
0.840 	& 70.66 \\
0.710	& 61.49 \\
0.600	& 56.29 \\
0.500	& 46.38 \\
0.420	& 42.16 \\
0.350 	& 36.41 \\
0.297	& 30.50 \\
0.250 	& 28.30 \\
0.210 	& 25.85 \\
0.177 	& 22.58 \\
0.149	& 20.49 \\
0.105	& 16.39 \\
0.088	& 15.53 \\
0.075	& 14.86 \\
0.050	& 8.34 \\
0.036	& 8.34 \\
0.025	& 6.95 \\
0.017	& 6.25 \\
0.013	& 6.25 \\
0.009	& 3.46 \\
0.006	& 2.76 \\
0.005	& 1.92 \\
0.003	& 1.09 \\
0.002	& 0.00 \\\hline\hline
\end{tabular}
\end{minipage}
\hspace{2em}
\begin{minipage}[c]{.6\linewidth}
\begin{center}
\includegraphics[angle=-90,width=.9\linewidth]{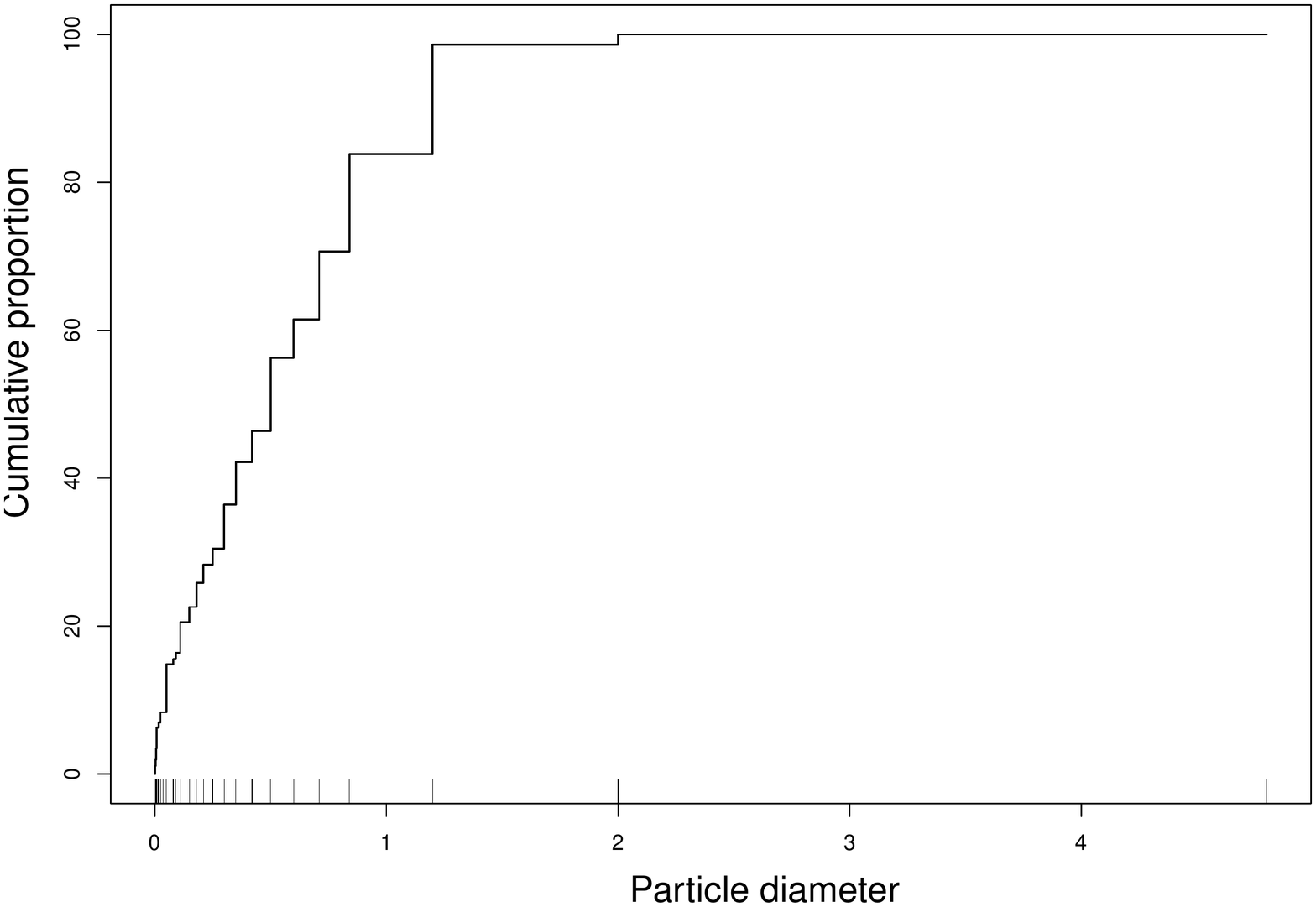}\\
(a) Original granulometric curve
\includegraphics[angle=-90,width=.9\linewidth]{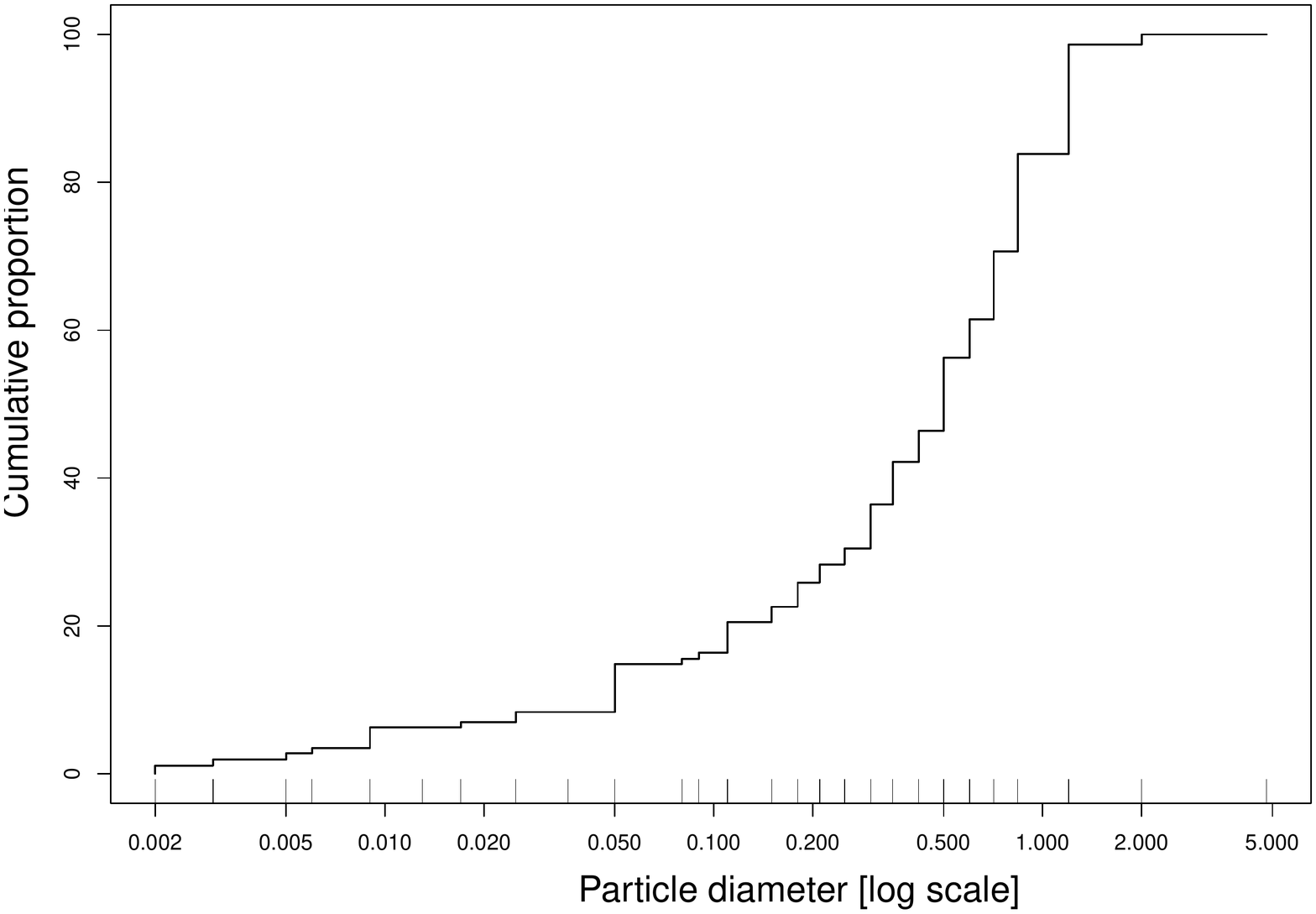}\\
(b) Granulometric curve in semilogarithmic scale
\end{center}
\end{minipage}
\caption{Granulometric data and curves: linear (a, top) and semilogarithmic (b, bottom), sample 1}\label{c1}
\end{figure}

The samples used as input for this analysis are presented in Table~\ref{tab:soiltype}.
Figure~\ref{c1} presents the result of the sieving process applied to Sample~1.
The table shows the diameters, in decreasing order and in mm, and the corresponding cumulative percentage of passing particles.
This same information is shown in the form of a granulometric curve in linear scale (right top); the sieves diameters are show superimposed to the abscissa axis.
The clutter effect of small sieves is evident, therefore the frequent choice of presenting such curves in semilogarithmic scale (right bottom).
Notice that the diameters are not evenly distributed deserving, thus, a careful treatment.

\begin{table}[hbt]
\begin{center}
\caption{Soil types analized}\label{tab:soiltype}
\begin{tabular}{r  l r}
\toprule
Sample & Soil type & Porosity $\eta$\\ \midrule
1 & sand, silt with traces of de gravel & 0.35\\
2 & sand with traces of silt & 0.43\\
3 & sand, portions of clay, traces of silt and gravel & 0.42\\
4 & sand with traces of silt & 0.39\\
5 & sandy clay with portions of silt & 0.40\\
6 & sandy clay with portions of silt and traces of gravel & 0.33\\
7 & sand & 0.44\\ \bottomrule
\end{tabular}
\end{center}
\end{table}

\begin{figure}
\centering
\includegraphics[angle=-90,width=1\linewidth]{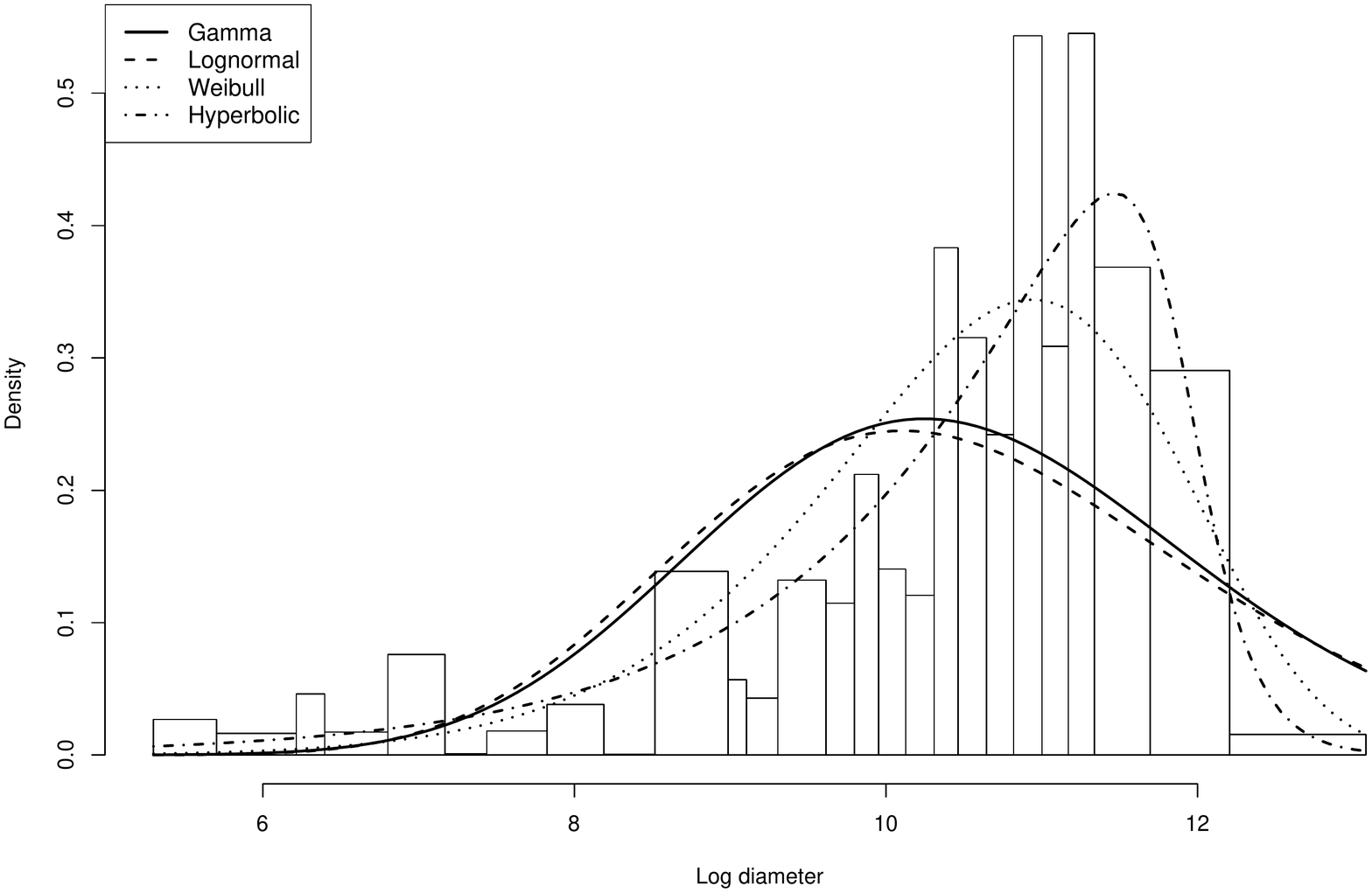}
\caption{Loghistogram and fitted density for sample 1 }\label{histogram1}
\end{figure}

\begin{figure}[hbt]
\centering
\subfigure[Sample 2\label{c3}]{\includegraphics[angle=-90,width=.49\linewidth]{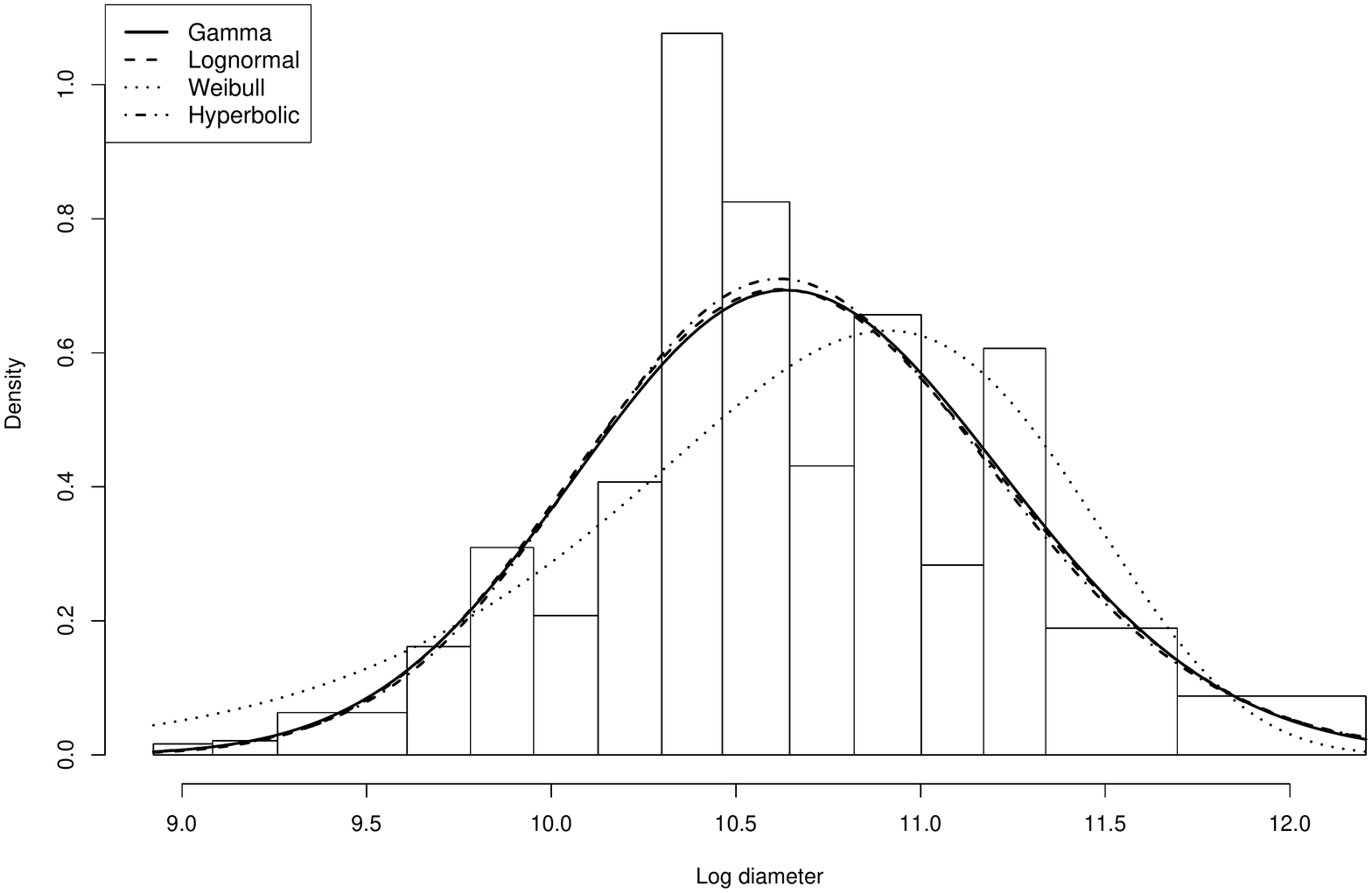}}
\subfigure[Sample 3\label{c5}]{\includegraphics[angle=-90,width=.49\linewidth]{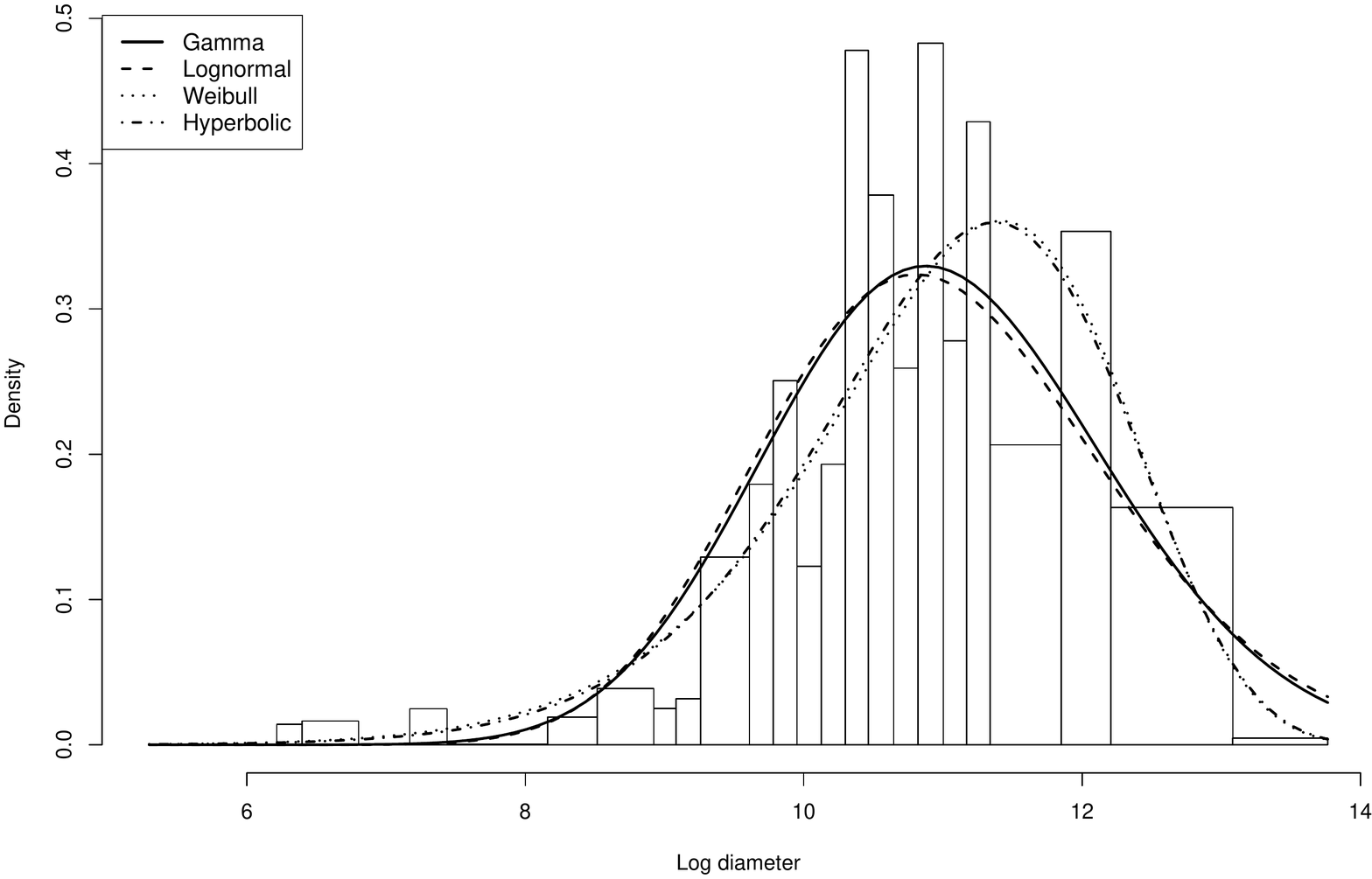}}
\subfigure[Sample 4\label{c6}]{\includegraphics[angle=-90,width=.49\linewidth]{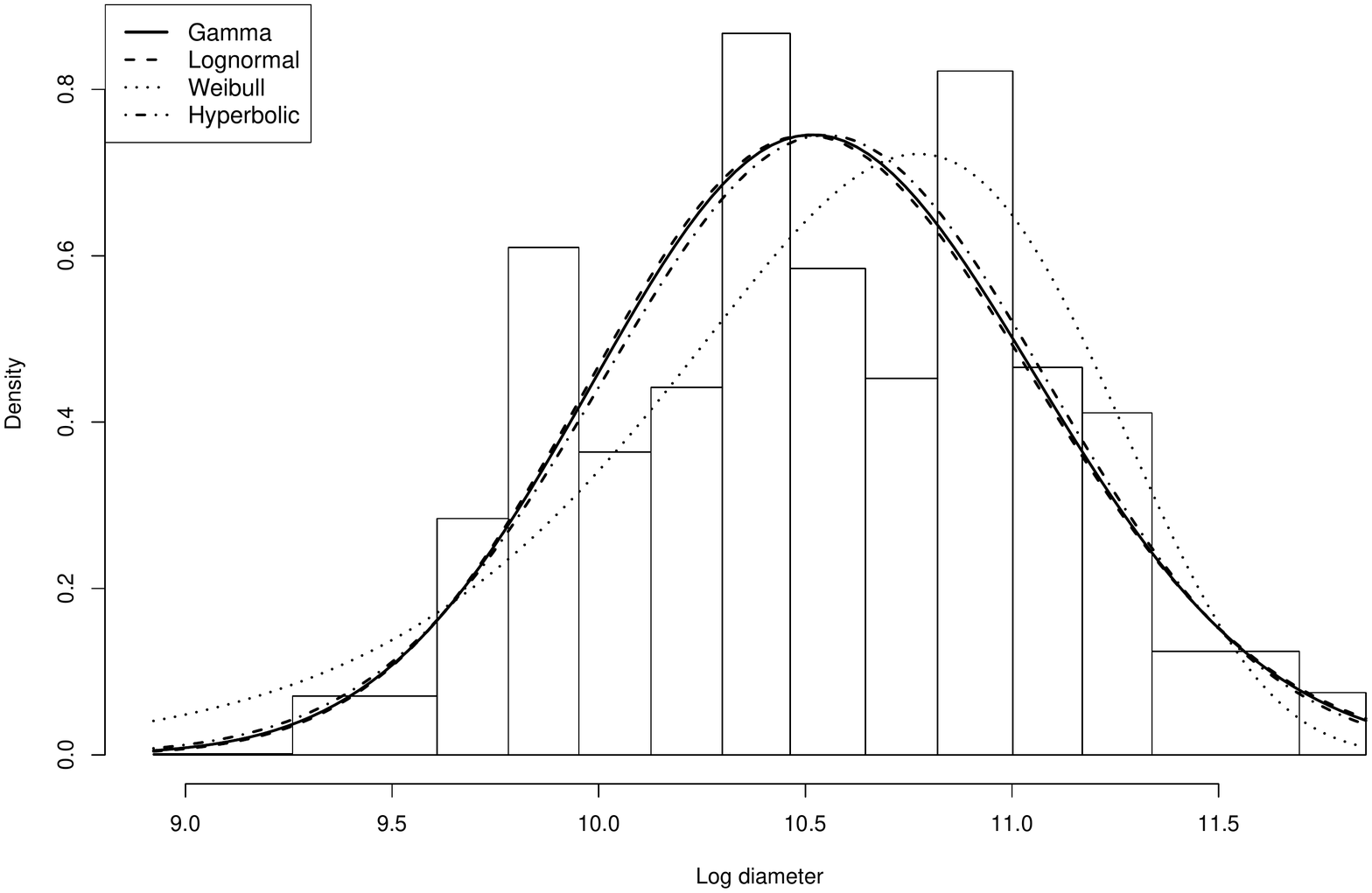}}
\subfigure[Sample 5\label{c7}]{\includegraphics[angle=-90,width=.49\linewidth]{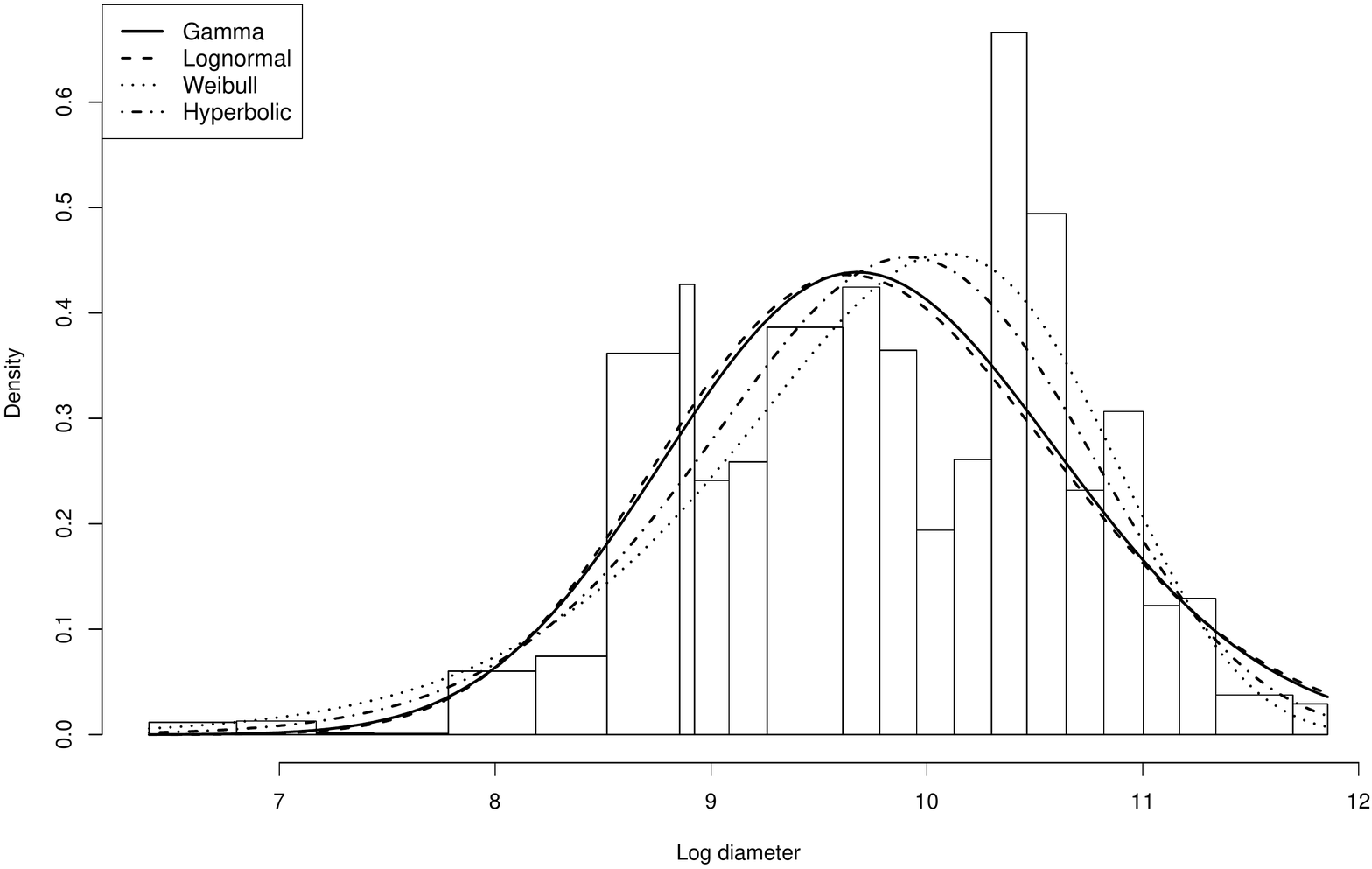}} 
\subfigure[Sample 6\label{c9}]{\includegraphics[angle=-90,width=.49\linewidth]{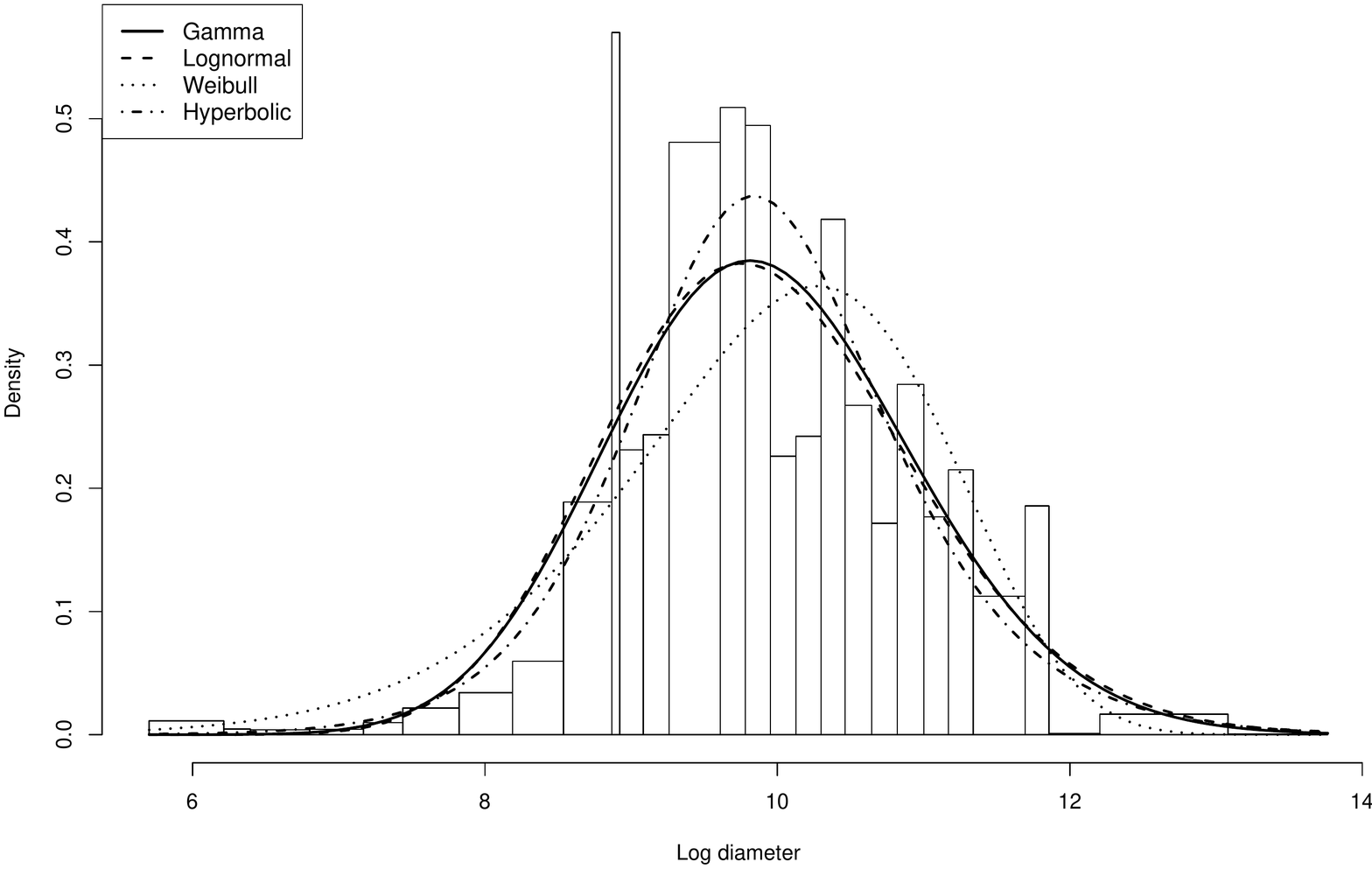}}
\subfigure[Sample 7\label{c10}]{\includegraphics[angle=-90,width=.49\linewidth]{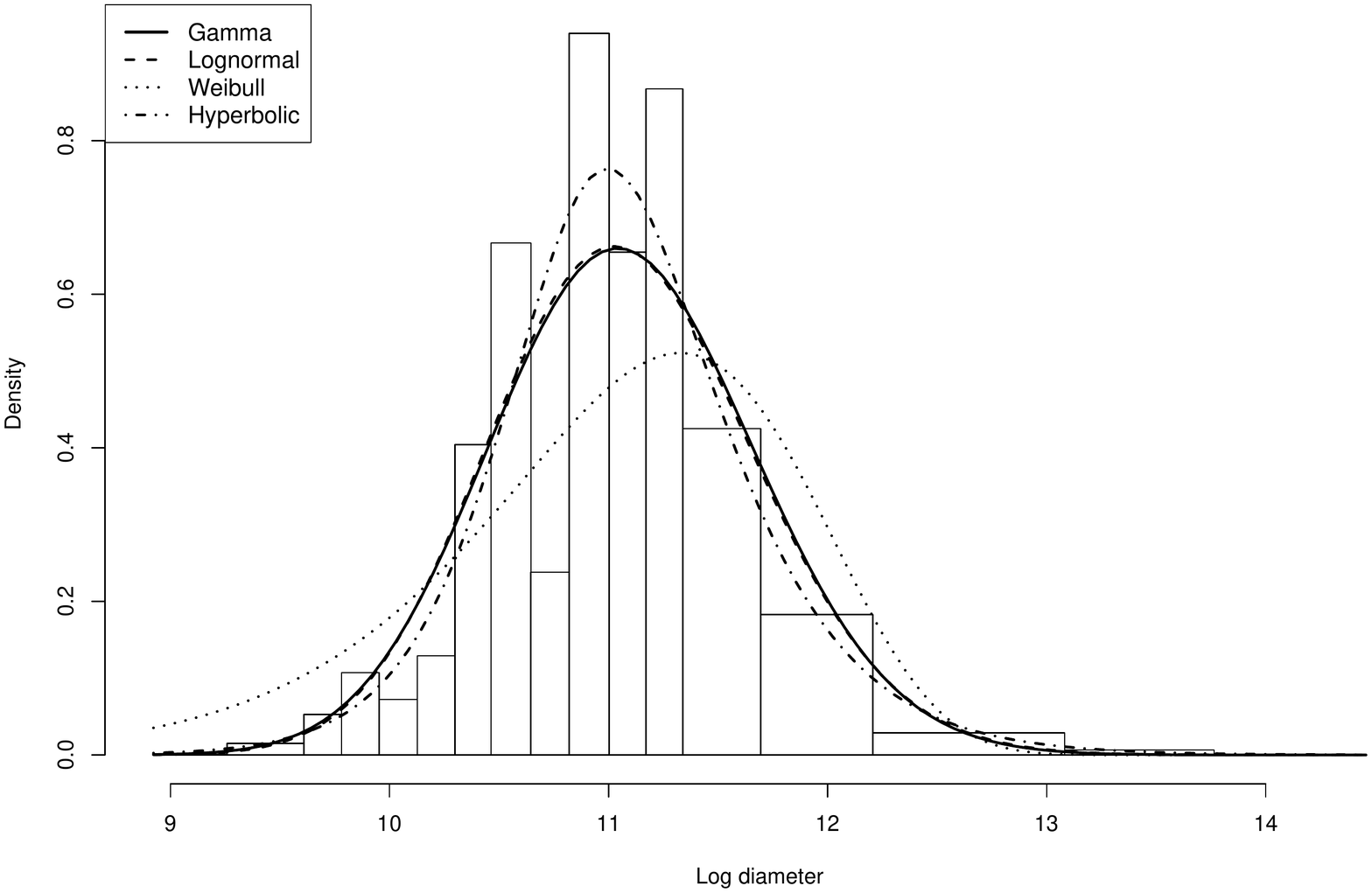}}
\caption{Loghistograms and fitted densities}\label{histograms}
\end{figure}

Each data set from Table~\ref{tab:soiltype} was analyzed using the procedure presented in previous sections, i.e., (i)~its granulometric curve was obtained by sieving (see Figure~\ref{c1} as an example), (ii)~a histogram was formed (Figures~\ref{histogram1} and~\ref{histograms}), (iii)~samples of pseudorandom diameters were drawn.
These diameters were then analyzed by fitting the gamma, lognormal, Weibull and hyperbolic distributions.
Maximum likelihood estimation was performed using the \verb+fitdistr+ function from the \verb+MASS+ package of \texttt R (for the gamma, lognormal and Weibull distribution) and the \verb+hyperbFit+ function from the \verb+HyperbolicDist+ package (for the hyperbolic distribution).
The estimated parameters, rounded to the third decimal place, are given in Table~\ref{estimatedparameters}.

\begin{table}[hbt]
\caption{Estimated parameters}\label{estimatedparameters}
\centering
\begin{tabular}{rrrrrrrrrrr}
\toprule
\multirow{2}{*}{\rotatebox{90}{Sample}} & \multicolumn{2}{c}{\textbf{Gamma}} & \multicolumn{2}{c}{\textbf{Lognormal}} & \multicolumn{2}{c}{\textbf{Weibull}} & \multicolumn{4}{c}{\textbf{Hyperbolic}}\\
\cmidrule(r){2-3} \cmidrule(r){4-5} \cmidrule(r){6-7} \cmidrule(r){8-11}
& $\widehat\alpha$ & $\widehat\beta$ & $\widehat\mu$ & $\widehat\sigma$ & $\widehat\alpha$ & $\widehat\lambda$ & $\widehat\pi$ & $\widehat\zeta$ & $\widehat\delta$ & $\widehat\mu$ \\ 
\midrule
1 & $43.717$ & $4.170$ & $2.338$ & $0.159$ & $10.281$ & $11.046$ & $-1.055$ & $0.750$ & $0.403$ & $11.892$ \\
2 & $342.999$ & $32.155$ & $2.366$ & $0.054$ & $18.806$ & $10.943$ & $0.235$ & $15.073$ & $ 2.072$ & $10.132$ \\
3 & $81.896$ & $7.437$ & $2.393$ & $0.113$ & $11.241$ & $11.513$ & $-3.369$ & $19.473$ & $1.367$ & $15.977$ \\
4 & $387.377$ & $36.738$ & $2.354$ & $0.051$ & $21.186$ & $10.799$ & $-0.138$ & $705.286$ & $14.081$ & $12.495$ \\
5 & $114.429$ & $11.722$ & $2.274$ & $0.095$ & $12.564$ & $10.163$ & $-1.171$ & $45.259$ & $3.816$ & $14.381$ \\
6 & $90.818$ & $9.153$ & $2.289$ & $0.106$ & $10.231$ & $10.381$ & $0.107$ & $2.635$ & $1.304$ & $9.698$ \\
7 & $334.601$ & $130.210$ & $2.403$ & $0.055$ & $16.146$ & $11.372$ & $0.159$ & $2.025$ & $0.631$ & $10.895$ \\
\bottomrule
\end{tabular}
\end{table}

The $p$-values of the $\chi^2$ test were computed, and they all resulted above $0.99$ with one exception, the Weibull distribution for sample~7.
Therefore, for the samples here analyzed, the choice of the ``best'' model can be guided by the computational cost of producing pseudorandom deviates or by any other criterion.

\begin{figure}[hbt]
\subfigure[Complete simulation\label{fig:CompleteSample9}]{ \includegraphics[width=.6\linewidth]{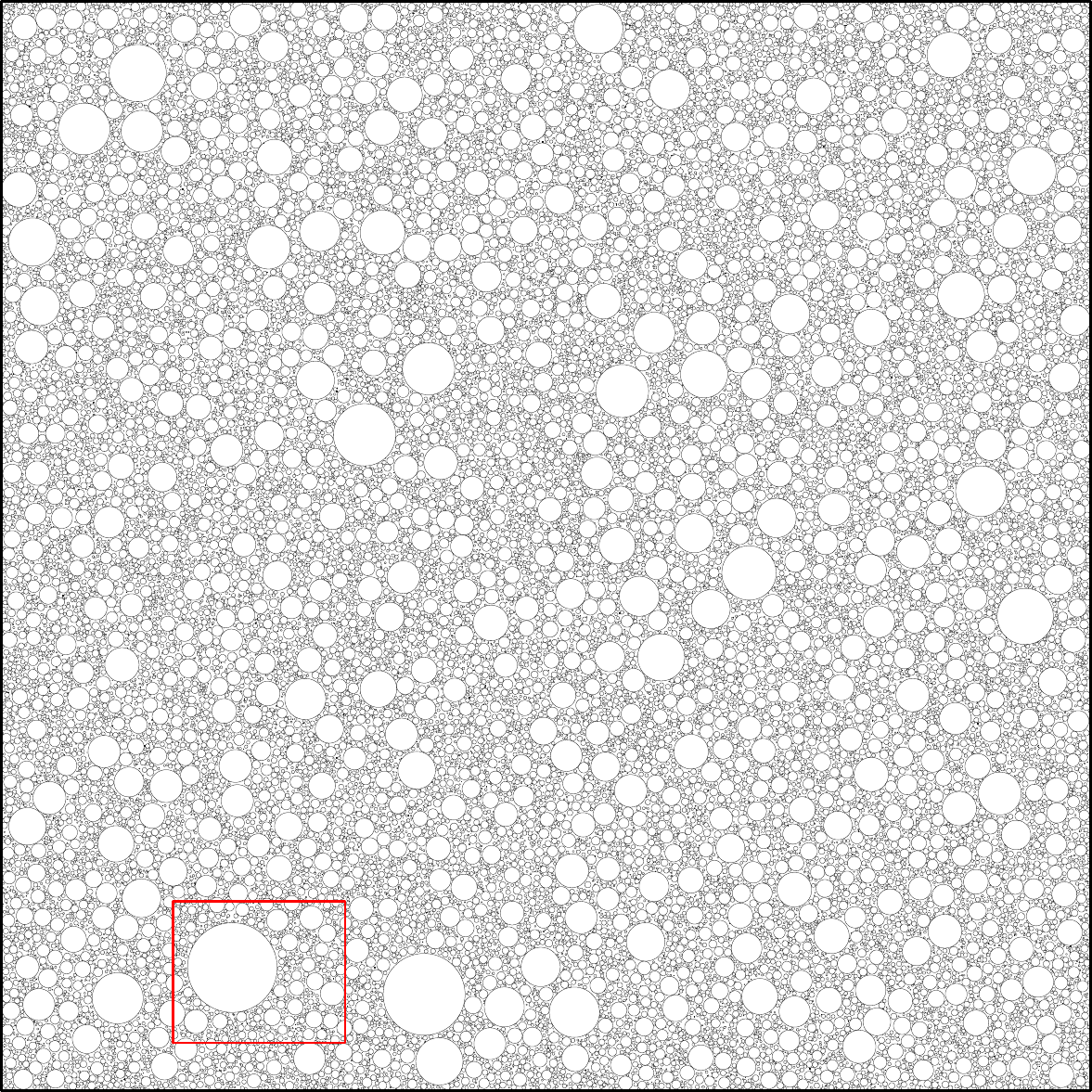}}
\subfigure[Detail\label{fig:DetailSample9}]{ \includegraphics[width=.38\linewidth]{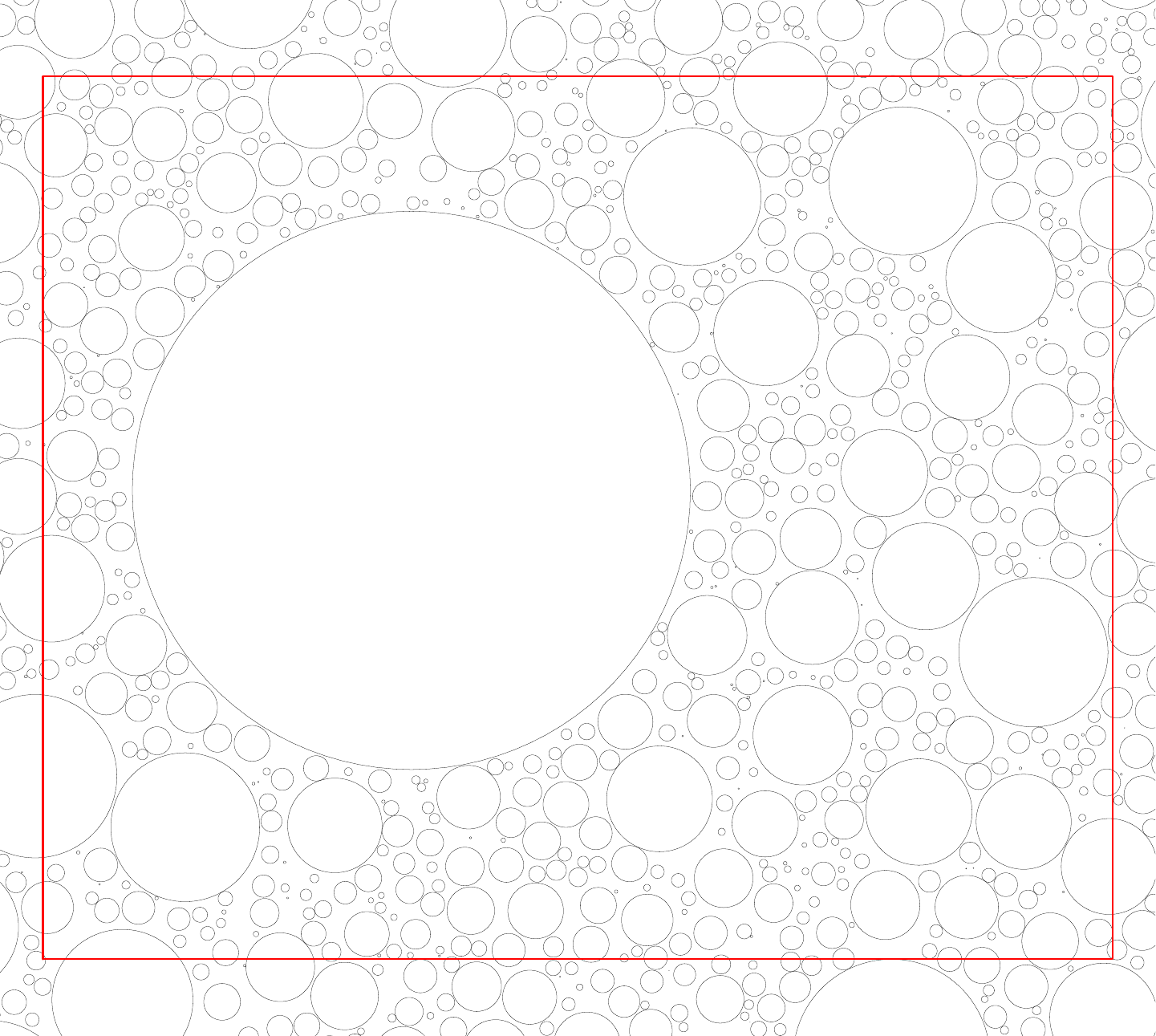}}
\caption{Simulated soil from Sample 6}\label{fig:SimulSample9}
\end{figure}

We simulated samples from each of the types of soil previously analized using the models presented in Table~\ref{tab:models}, which describe the logarithm of the diameters.
These simulations were performed in squared boxes specifying the desired porosity.
Table~\ref{tab:models} also presents the number of particles generated in each case ($N$), along with the desired and obtained porosities ($\eta$ and $\widehat\eta$, respectively).
Given the similarity of the results, in Figure~\ref{fig:SimulSample9} we only present the simulation corresponding to Sample~6.
Figure~\ref{fig:CompleteSample9} presents the $34688$ particles that were generated to fill a box with the specified porosity, while Figure~\ref{fig:DetailSample9} shows a detail where both the varibility of the model and the non-overlapping results are enhanced.

\begin{table}[htb]
\caption{Chosen model for each sample, number of particles, desired and obtained porosities}\label{tab:models}
\centering
 \begin{tabular}{rcccc}\toprule
  Sample & $\mathcal D$ & $N$ & $\eta$ & $\widehat\eta$ \\ \midrule
1 & Weibull & $6704$ & $0.350$ & $0.345$ \\
2 & Hyperbolic & $4846$ & $0.430$ & $0.421$ \\
3 & Gamma & $1922$ & $0.420$ & $0.418$ \\
4 & Lognormal & $7840$ & $0.390$ & $0.390$ \\
5 & Weibull & $27486$ & $0.400$ & $0.399$ \\
6 & Weibull & $34688$ & $0.330$ & $0.328$ \\
7 & Hyperbolic & $1611$ & $0.440$ & $0.430$ \\ \bottomrule
 \end{tabular}
\end{table}

\section{Conclusions and future work}\label{sec:conclusion}

We proposed and assessed a fast geometric algorithm which produces spatially homogeneous samples with specified porosity and granulometric curve using disks.
Starting with the granulometric curve and porosity of a soil, our algorithm produces samples of any size in closed arbitrary domains.
The outcomes of this algorithm can be used as granular media samples in computational simulations, e.g., discrete elements.

The algorithm allows the use of any suitable distribution for the particle size.
It was assessed with seven samples and four distributions, and in every case it produced samples wich mimic well the input data.

The proposed strategy can also be used in the simulation of other physical systems.
Kadau et al.~\cite{Kadau2009} propose a two‐dimensional contact dynamics model as a microscopic description of a collapsing suspension/soil to capture the essential physical processes underlying the dynamics of generation and collapse of the system.
Kadau and Herrmann~\cite{Kadau2011} study the influence of the granular Bond number on the
density profiles and the generation process of packings, generated by ballistic deposition under gravity.
Both works deal with loose granular media and, thus, there is no need to impose strong physical constraints.

The research continues with the proposal of a hybrid technique which, starting with our geometrical algorithm, produces a final configuration in geostatic equilibrium by relaxation in both 2D and 3D domains.

\begin{acknowledgements}
The authors are grateful to CNPq, Fapeal, Finep and Petrobras.
The anonymous referees provided enlightening comments and suggestions. 
\end{acknowledgements}

\bibliographystyle{spmpsci}
\bibliography{ref}

\end{document}